\documentclass[onecolumn,amsfonts,amssymb,amsmath,floatfix]{article}
\usepackage[dvips]{graphicx}
\usepackage{natbib}
\usepackage{bm}
\bibpunct{(}{)}{;}{a}{}{,}
\title{Scalar potential model progress}

\author{J.C. Hodge$^{1}$\thanks{E-mail:jch9496@blueridge.edu}\thanks{Visiting from XZD Corp., 16 Hosta Ln., Brevard, NC, 28712, E-mail:{scjh@citcom.net}}\\
$^{1}${Blue Ridge Community College, 100 College Dr., Flat Rock, NC, 28731-1690}}

\begin{document}

\maketitle

\begin{abstract}
Because observations of galaxies and clusters have been found inconsistent with General Relativity (GR), the focus of effort in developing a Scalar Potential Model (SPM) has been on the examination of galaxies and clusters.  The SPM has been found to be consistent with cluster cellular structure, the flow of IGM from spiral galaxies to elliptical galaxies, intergalactic redshift without an expanding universe, discrete redshift, rotation curve (RC) data without dark matter, asymmetric RCs, galaxy central mass, galaxy central velocity dispersion, and the Pioneer Anomaly.  In addition, the SPM suggests a model of past expansion, past contraction, and current expansion of the universe.  GR corresponds to the SPM in the limit in which the effect of the Sources and Sinks approximate a flat scalar potential field such as between clusters and on the solar system scale, which is small relative to the distance to a Source.  
\end{abstract}
%\begin{keyword}
% keywords here, in the form keyword \sep keyword
% PACS code here, in the form \PACS code \sep code
galaxies:distances -- redshifts -- galaxies:spiral -- cosmology:theory 
%\end{keyword}
%\end{frontmatter}

% main text
\section{INTRODUCTION}

Explaining and predicting observations that are unsatisfactorily accounted by the cosmological standard model and by general relativity (GR) motivates the investigation of alternate models.  Examples of these future model challenges are (1) the need for and the ad hoc introduction of dark matter and dark energy, (2) the Pioneer Anomaly (PA), (3) the incompatibility of GR with observations of subatomic particles, (5) the need for fine - tuning of significant parameters, (6) the incompatibility of GR with galactic scale and galaxy cluster scale observations~\citep{sell}, (7) the poor correlation of galactocentric redshift $z$ of galaxies to distances $D$ (Mpc) $>10$ Mpc measured using Cepheid stars \citep[see data in]{free,macr,saha}, (8) the lack of a galaxy and galaxy cluster evolution model consistent with observations, and (9) the lack of application of Mach's Principle.  Other examples \citep{arp,peck} that proponents of the standard model dispute the interpretation of the observational data include evidence (1) for discrete redshift (2) for QSO association with nearby galaxies, (3) that galaxies are forming by ejection from other galaxies rather than merging, and (4) that $z$ may be caused by a phenomena other than Doppler shift, called ``intrinsic'' $z$.

A new physics scalar potential model (SPM) has been proposed \citep{hodg765,hodg344}.  The SPM postulates a heat equation model with Sources, Sinks, and a non-adiabatic universe.

This Paper is a review of the SPM and a preliminary comparison of the SPM with GR Principles.  Section~\ref{sec:review} reviews the progress of the SPM.  Comments about the fine-tuning problems; Mach's Principle; the Equivalence Principle; and redshift and universe expansion are discussed in Sections~\ref{sec:fine}, \ref{sec:mach}, \ref{sec:ep}, and \ref{sec:redshift}, respectively.  The discussion and conclusion are in Sections~\ref{sec:disc}. 

\section{\label{sec:review}SPM review}

The SPM postulates the only fundamental components of the universe are the ``plenum''\footnote{Taken from Aristotle's model of the stuff between matter rather than Democritus's model of the ``void'' between atoms.  Further, the ``aether'' model of the 19$^\mathrm{th}$ century is inappropriate because of the implication that the aether obeyed a wave equation rather than the heat equation, because the aether model did not include Sources and Sinks, and because the aether was considered a universal rest frame.} and matter in three-dimensional (3D) Euclidean space.  The scalar potential $\rho$ is the amount of the plenum per volume.  The 3D space of the SPM is a passive backdrop.  Euclidean distance is measured by timing events such as using Cepheid stars rather than using observations of matter that travels through the plenum such as $z$ measurement.  The plenum is ubiquitous, has a corporeal existence, and exerts a positive-pressure.  Like Space\footnote{The upper case ``S'' is to distinguish the GR Space that is influenced by matter and influences matter from the passive backdrop space (lower case ``s'') of the SPM.} of GR, matter acts to shape the plenum and the plenum's $\vec{\nabla} \rho$ is proportional to a force that acts on matter.  The effort has been to discover the characteristics and interaction of the fundamental components.  Attention has been focused on observations that are inconsistent with or are poorly described by the standard model.  

The plenum and matter must interact.  Spiral galaxies are postulated to be Sources of the plenum and of matter.  Elliptical and other galaxies are postulated to be the Sinks of the plenum and of matter.  Thus, the plenum flows and the universe is not adiabatic.  The heat equation applies to the plenum.  

If matter and the plenum interact and if light is composed of matter, then the $z$ of photons depends on the $\rho$ field between the emitter and receiver.  Following fluid and optics analogies, the $z$ is primarily caused by the amount, density, and turbulence (inhomogeneity) of the $\rho$ field.  The application of the SPM improved the $D$ -- $z$ correlation from less than 0.80 to 0.88 \citep{hodg344}.

Because the intergalactic $\rho$ field shape derives from Sources and Sinks, the proximity of galaxies to the light path influences $z$.  Thus, the discrete $z$ observations are a genuine effect \citep{hodg344}.  

Observations suggest galaxy clusters are structured as cells \citep{hodg344}.  The IGM observations suggest matter flows from spiral galaxies to elliptical galaxies.  The term ``galaxy cluster'' is sometimes applied to any bunch of galaxies that are close together from our viewpoint.  Because galactic distance is a factor in determining ``close'', a galaxy cluster is model dependent.  For example, that the QSO's with high $z$ and angularly near galaxies with lower $z$ may be close is disputed.  The SPM concept of a ``cell'' is a galaxy cluster with Sources and Sinks that are distant from other cells and wherein all matter flows within the cell.  A cell may be as small as one Source and one Sink.  An analogy to fluid dynamics suggests the cell structure is like a Rankine Oval.  

The cell structure of galaxy clusters and observations of elliptical galaxies suggest the Sink rate of matter is dependent on the amount of matter in the Sink galaxy.  Therefore, a feedback mechanism analogous to the thermal control of a thermostat is suggested.  The temperature $T$ of the universe ``hunts'' 2.718 K and its spectrum should be black body radiation \citep{hodg140}.  Thus, the horizon problem is explained.  If \mbox{$T>2.718$ K}, the plenum, but not matter, is flowing out of the Rankine Oval of cells. 

The structure of spiral galaxies suggests a Source is at the center of spiral galaxies.  A galaxy's B-band luminosity $L_\mathrm{B}$ was found to correlate to the Source or Sink strength in the galaxy \citep{hodg029}.  Therefore, the $ L_\mathrm{B}$ is correlated with $z$ and with spiral galaxy observations of rotation curves (RCs), of central mass, of central velocity dispersion, and of asymmetric RCs \citep{hodg029,hodg699}.  The $\vec{\nabla} \rho$ from neighbor galaxies was found to correlate to rotation curve asymmetry and was found to be a correction term to the RC equations \citep{hodg029}.  The spiral galaxy observations require the $\vec{\nabla} \rho$ field to be repulsive of matter and to act on a property of matter $m_\mathrm{s}$ other than its gravitational mass $m_\mathrm{g}$.  The SPM suggests $m_\mathrm{s}$ is the cross-section of particles.  The HI RCs; the differing HI and H$_\alpha$ RCs; and metallicity-radius relation suggest the largest particle (LP) on which the $\rho$ field acts as a single particle is the atom.  That is, if one such particle is directly behind another such particle, the front particle masks the back particle and the total $m_\mathrm{s}$ is the cross-section of only the front particle.  Conversely, matter acts on the $\rho$ field in proportion to the 3D bulk property of $m_\mathrm{g}$.

The apparent paradox of a spiral galaxy's central parameters being related to the galaxy's global properties through $L_\mathrm{B}$ and to $z$ is resolved if the galaxy matter has no effect on intergalactic scales.  The plenum flows around matter such that the $\rho$ field far downstream is as if the matter was not present.  Therefore, the analogy of fluid flow is reinforced \citep{hodg567}. 

The SPM of $z$ allows a blueshift (energy gain) as well as a redshift (energy loss).  On the intergalactic scale, in the absence of galaxies along the light path, the Hubble Law results from $\rho \propto R^{-1}$, where $R$ is the radial distance between points. 

In addition to the Sun directed blueshift, there are other characteristics of the PA \citep{ande02} that a model of the PA should explain.  The PA has an apparent annual periodicity.  Although within uncertainty limits, the Pioneer 11 (P11) spacecraft anomaly \mbox{$a_\mathrm{P11}$} may be slightly larger than the Pioneer 10 (P10) anomaly \mbox{$a_\mathrm{P10}$}.  The sunward acceleration $a_\mathrm{P}$ calculation by the {\textit{Sigma}} and CHASMP program methods for P10 (I) and P10 (II) show a discrepancy while showing consistency for P10 (III) \citep[Table I]{ande02}.  The $a_\mathrm{P}$ of both spacecraft may be declining with distance \citep[as shown by the envelope in Fig.~1]{tury}.  The blue shift of the PA is significantly smaller \citep{ande02,niet05a} immediately before P11's Saturn encounter.  The value of $a_\mathrm{P}$ averaged over a period during and after the Saturn encounter had a relatively high uncertainty \citep{niet05a}.  That $a_\mathrm{P} \approx c H_\mathrm{o}$, where $c$ (cm s$^{-1}$) is the speed of light and $H_\mathrm{o}$ (s$^{-1}$) is the Hubble constant, suggest a cosmological connection to PA.  The PA has an apparent sidereal day periodicity. 

Several new physics models have been proposed \citep{ande02,bert}.  \citet{bert} concluded a scalar field is able to explain the PA.  The observation of the PA suggests matter curves the $\rho \propto -R^{-1}$ as also suggested by GR's gravity effect on Space.  The SPM of the PA postulates gravity is the action of the $\vec{\nabla} \rho$ field.  The same equation used to calculate the $D$ -- $z$ relationship was applied to the PA.  To date, only the SPM explains all the noted data of the PA including the cosmological connection without an impact on the planetary ephemerides \citep{hodg567}.  

The other proposed models of the PA are inconsistent with the PA data \citep{ande02,toth}, with the planetary ephemerides, or with the Weak Equivalence Principle (WEP).  Further, the other proposed models attempt to explain only the general value of $a_\mathrm{P}$ and ignore the other PA characteristics.  The PA exceeds by at least two orders of magnitude the GR corrections to Newtonian motion.  Cosmic dynamics according to GR has far too little influence in galaxies to be measurable and expansion of the universe is negligible for scales up to galactic clusters \citep{coop,sell}.  Further, the expansion of the universe indicated by $z$ has a sign opposite to $a_\mathrm{P}$. 

Therefore, gravity is caused by matter and by the Source-Sink flow.  The forces interact by contact rather than the ad hoc introduction of ``action at a distance'' or of an elusive graviton.  The $\vec{\nabla} \rho$ is directed away from cells and from spiral galaxies.  The $\vec{\nabla} \rho$ is directed toward Sinks and matter.  Therefore, the SPM plenum force on matter is a broader model of gravity rather than a ``fifth force''.

If the $\rho$ field flowing from Sources to Sinks is relatively flat, $\vec{\nabla} \rho$ is static (non-flowing) and is due to only matter.  In such a $\rho$ field the flow of the plenum is small although $\vec{\nabla} \rho$ is not.  The mass effect on a non-flowing plenum is extended farther than in a flowing, high $\vec{\nabla} \rho$ field.  GR is consistent with data from our solar system, in which the distance across the system is small relative to the distance to a Source, and may be consistent with data from intercluster scales, which are outside the Rankine Oval effect.  Over distances where the plenum flow is significant, GR is inconsistent with the observations.  Therefore, GR corresponds to the SPM in the limit in which the effect of the Sources and Sinks approximate a flat $\rho$ field.  

The SPM development and confrontation with observations suggest several new or altered fundamental postulates or Principles.

\section{\label{sec:fine}Fine-tuning problems}

The SPM proposes parameters that have a fine-tuning characteristic such as the coincidence problem and the flatness problem and parameters that apparently have the same value throughout the universe such as the horizon problem are a result of a negative feedback condition.  Thus, the $T$ ``hunts'' 2.718 K in each cell, the $L_\mathrm{B}$-central mass balance is maintained, and $L_\mathrm{B}$-RC balance is maintained.  Feedback is a well-understood concept in engineering as a fine-control mechanism.  

\section{\label{sec:mach}Mach's Principle}

Ernst Mach suggested the gravitation of matter had its origin in the distribution of matter in the universe.  In the SPM, Mach's Principle is manifest by the dependence of the $\rho$ field on the strength and position of Sources, Sinks, and matter.  Because the plenum flows, the plenum is continually changing rather than being a universal rest frame.  In GR the energy-momentum tensor is transformed into the geometry of Space.  In the SPM the energy and momentum are defined by the $\rho$ field. 

\section{\label{sec:ep}Equivalence Principle}

The concept of two forms of energy was born in Galileo's free fall experiments.  The interaction of the plenum and matter produces two forms of energy.  One depends on the action of the $\rho$ field on matter that is kinetic energy.  The other is the effect of matter on the $\rho$ field that produces the potential energy field.  Because $T \approx 2.718$ K, the relative Source and Sink strengths are balanced by a feedback mechanism to maintain the potential and kinetic energy balance.  Thus, the resolution of the flatness problem is a natural consequence of the SPM. 

The GR WEP implies that a kinetic property measured by acceleration and a charge-type property measured by gravitational potential are equal \citep{unzi}.  In GR a type of redshift dependent on the gravity potential field ($G M /R$) derives from the WEP.  If the emitter and observer are in a uniform gravitational field $\propto -R^{-1}$, the electromagnetic EM signal acquires or looses energy because of the change in gravitational field potential.  The Pound-Rebka experiment of 1959-1960 measured the WEP redshift \citep{pound}\footnote{This should not be confused with the theorized GR time dilation effect ($1+z$) caused by the differing gravitation fields of emitter and observer. }.  The WEP suggests the integral of the EM frequency change experienced at each small increment along the path could be used to calculate $a_\mathrm{P}$.  The SPM uses the gravitational potential from all matter at each point along the EM signal's path to calculate $a_\mathrm{P}$.  The difference between GR and the SPM is that the SPM also considers the inhomogeneity of the gravitational field in the $K_\mathrm{i} I$ term and the amount of plenum the EM signal passes through in the $K_\mathrm{dp} D_\mathrm{l} P$ term.  Without the $K_\mathrm{i} I$ term or $K_\mathrm{dp} D_\mathrm{l} P$ term in the calculation, the correlation of measured and calculated $a_\mathrm{P}$ would be much poorer to the degree the WEP method fails.

The SPM postulates matter experiences a force $\vec{F} = G_\mathrm{s} m_\mathrm{s} \vec{\nabla} \rho$.  In the case of a $\rho$ field variation produced solely by matter, 
\begin{equation}
F=\left(G_\mathrm{s} m_\mathrm{s} \right) \left(\frac{ G_\mathrm{g} M_\mathrm{g}}{R} \right)
\label{eq:1},
\end{equation}
where $G_\mathrm{g}$ and $M_\mathrm{g}$ are the proportionality constant and bulk property of matter of the upper case particle $M$ called mass, respectively, and $G_\mathrm{s}$ is the proportionality constant of $m_\mathrm{s}$ property of lower case particle $m$ of matter.  
The $ G_\mathrm{g} M_\mathrm{g}/R$ factor is interpreted as characterizing the acceleration field.  

For an assembly of gravitationally bound particles, the total $m_\mathrm{s}$ is the sum of the total number of the LP upon which the $\rho$ field acts as individual particles.  

The semi empirical mass for the internal energy $E^A$ generated for atomic weight $Z$ and atomic number $A$ indicates a gravitational effect for differing nuclei \citep[Equations 2.5 to 2.12]{will} and is thought to place an upper limit on the WEP violation if any.  Further, isotopes with the same $Z$ and different $A$ have differing $E^A$ values.  In the experiments performed to examine such factors, the bulk property of the masses ($m_\mathrm{g}$ and $M_\mathrm{g}$) are used in the calculations.  Equation~(\ref{eq:1}) implies the WEP test should be done with the same attractor and with the same number of atoms in the pendulum or free-fall body for the differing isotopes.  Maintaining equal $m_\mathrm{g}$ considers only the bulk property of matter that reduces the SPM effect.

Because $F$ is symmetric, \mbox{$(G_\mathrm{s} m_\mathrm{s}) (G_\mathrm{g} M_\mathrm{g}) = (G_\mathrm{g} m_\mathrm{g}) (G_\mathrm{s} M_\mathrm{s}) = G m_\mathrm{g} M_\mathrm{s}$}.  If the LP is quarks, the $(G_\mathrm{g} /G_\mathrm{s})$ ratio indicates the relative volume to cross-section that may differ for quarks and larger structures.  For example, if the LP are quarks, differing $A/Z$ ratios with equal $Z$ will determine differing $G$ values.  For elements the number of protons and the number of neutrons are approximately equal.  Therefore, the $G$ varies little among atoms.

However, the WEP deals with inertial mass $m_\mathrm{i}$ rather that $m_\mathrm{s}$.  Therefore, the SPM suggests a test of the WEP different from those done previously is required to differentiate the three mass interaction parameters $m_\mathrm{i}$, $m_\mathrm{g}$, and $m_\mathrm{s}$.  

\section{\label{sec:redshift}Redshift and universe expansion}

In the decade after Hubble discovered the apparent magnitude -- $z$ relation, two types of $z$ models suggested were that the $z$ is a result of the Doppler effect and that $z$ is a result of photons loosing energy as they moved through space (tired light models).  The Doppler model required an expanding universe.  The SPM allows an energy gain from the plenum, which is observed in the blueshift of some galaxies and of the PA.  Therefore, the SPM of $z$ is an energy gain and loss model rather than a tired light model. 

The Big Bang (BB) cosmological models suggest the observed redshift of light from galaxies is a result of the Doppler effect.  Therefore, BB concludes, the Hubble constant is at the heart of cosmology and the universe of today evolved from a very dense core.  The SPM allows the Doppler shift from galaxy rotation in the cell to partially explain the $z$.  However, the primary factor determining $z$ is the characteristics of the $\rho$ field.  The Hubble constant is much less important and merely reflects an approximate $D$--$z$ relation.  The Doppler shift caused by the relative movement of the target galaxy and the Galaxy causes a deviation between measured redshift $z_\mathrm{m}$ and calculated redshift $z_\mathrm{c}$ caused by the $\rho$ field.  If there is no overall universal expansion such as in an adiabatic universe, the SPM expects $ <z_\mathrm{m} - z_\mathrm{c}>=0$, where ``$< \, >$'' means ``average''.  For a large sample that includes both approaching and receding galaxies in cells, the $<z_\mathrm{m} - z_\mathrm{c}>$ is a measure of universe expansion.  The SPM suggests the universe is currently expanding because $T>2.718$~K but at a much slower rate than suggested by the Hubble constant.  The ``hunting'' characteristic of the SPM \citep{hodg140} is currently consistent with a creation, cyclic cosmology and an eternal, cyclic universe.

\section{\label{sec:disc}Discussion and conclusion}

To be viable, a new cosmological model must at least describe the observations that are consistent with the currently popular standard model; be simpler or consistent with more types of observations; describe additional observations poorly or unsatisfactorily explained by the current model; and make predictions that differ from the current model and that are later confirmed.  A definite advantage for any new physics model would be if the link between the big (cosmology) and small (subatomic) were part of the model.  If the Hubble Law and $z$ do not reflect universe expansion, then the SPM has yet to confront the nucleosynthesis pillar and the CMB pillar of the standard model.  The SPM may include the nucleosynthesis of atoms and subatomic particles from the Source outward rather than from the evolution of the universe from the initial state.  Further, the SPM should develop a galaxy evolution model.

The SPM has been found to be consistent with galaxy cluster cellular structure, the flow of IGM from spiral galaxies to elliptical galaxies, intergalactic redshift without an expanding universe, discrete redshift, rotation curve (RC) data without dark matter, asymmetric RCs, galaxy central mass, galaxy central velocity dispersion, and the Pioneer Anomaly.  In addition, the SPM suggests a model of past expansion, past contraction, and current expansion of the universe.  Mach's Principle, the Equivalence Principle, GR time dilation redshift, redshift and universe expansion correspondence to the SPM was discussed.  GR corresponds to the SPM in the limit in which the effect of the Sources and Sinks approximate a flat $\rho$ field such as between clusters and on the solar system scale, which is small relative to the distance to a Source.  

%\acknowledgments
\section*{Acknowledgments}

I acknowledge and appreciate the financial support of Maynard Clark, Apollo Beach, Florida, while I was working on this project.

% The phrase \citep{Bai92} produces (Bailyn 1992).
% In the phrase \citeasnoun{Bai95} Bailyn et al. (1995) appear as a noun.
% Affixes (e.g. Barnes et al. 1976) are produced by the phrase
% \citeaffixed{Barnes et al. 1976}{e.g.}.
% Other options of the harvard package, e.g. \citeyear, are not
% reproduced in New Astronomy.

%\begin{thebibliography}{}

% \harvarditem{Name}{Year}{label}
% Text of bibliographic item
%\harvarditem{Author}{year}{label}Reference

%\end{thebibliography}

\end{document}